\documentclass[a4paper, 11pt]{article}
\usepackage{graphicx}\usepackage{amsmath}\usepackage{float}
\usepackage[inner=1in, top=1in, bottom=1in]{geometry}
\usepackage{amsfonts}
\usepackage{color,soul}
\usepackage{framed}
\usepackage{listings}
\usepackage{gensymb}
\usepackage{hyperref}
\usepackage{xcolor}
\usepackage{authblk}

\DeclareMathOperator{\Cov}{Cov}

\DeclareMathOperator{\Var}{Var}
\DeclareMathOperator{\E}{E}

\DeclareMathOperator{\x}{\boldsymbol{x}}
\DeclareMathOperator{\X}{\boldsymbol{X}}
\DeclareMathOperator{\y}{\boldsymbol{y}}

\DeclareMathOperator{\bmu}{\boldsymbol{\mu}}
\DeclareMathOperator{\bSigma}{\boldsymbol{\Sigma}}

\begin{document}

\title{Increasing the efficiency of Sequential Monte Carlo samplers through the use of approximately optimal L-kernels}


\author[1]{P. L. Green}
\author[2]{R. E. Moore}
\author[1]{R. J. Jackson}
\author[3]{J. Li}
\author[2]{S. Maskell}

\affil[1]{Department of Mechanical, Materials and Aerospace Engineering}
\affil[2]{Department of Electrical Engineering and Electronics}
\affil[3]{Department of Mathematical Science \\ University of Liverpool \\ Liverpool, UK \\ L69 7ZF}

\maketitle

\section*{Abstract}

By facilitating the generation of samples from arbitrary probability distributions, Markov Chain Monte Carlo (MCMC) is, arguably, \emph{the} tool for the evaluation of Bayesian inference problems that yield non-standard posterior distributions. In recent years, however, it has become apparent that Sequential Monte Carlo (SMC) samplers have the potential to outperform MCMC in a number of ways.  SMC samplers are better suited to highly parallel computing architectures and also feature various tuning parameters that are not available to MCMC.  One such parameter - the `L-kernel' - is a user-defined probability distribution that can be used to influence the efficiency of the sampler. In the current paper, the authors  explain how to derive  an expression for the L-kernel that minimises the variance of the estimates realised by an SMC sampler. Various approximation methods are then proposed to aid implementation of the proposed L-kernel. The improved performance of the resulting algorithm is demonstrated in multiple scenarios. For the examples shown in the current paper, the use of an approximately optimum L-kernel has reduced the variance of the SMC estimates by up to 99 \% while also reducing the number of times that resampling was required by between 65 \% and 70 \%. Python code and code tests accompanying this manuscript are available through the Github repository \url{https://github.com/plgreenLIRU/SMC_approx_optL}. \\

Key words: Sequential Monte Carlo, Markov chain Monte Carlo, Bayesian inference

\section{Introduction}

Markov Chain Monte Carlo (MCMC) is an approach through which, theoretically, it is possible to generate samples from an arbitrary probability distribution. MCMC has therefore become a key tool in the analysis of Bayesian inference problems as it can be used to generate samples from posterior probability distributions for which Bayes' theorem does not yield closed-form solutions. Beginning with the fundamental paper by Metropolis \textit{et al.} \cite{metropolis1953equation}, later generalised by Hastings \cite{hastings1970monte}, the popularity of MCMC has grown considerably. Myriad different MCMC approaches have now been proposed; algorithms such as Hamiltonian Monte Carlo (also known as Hybrid Monte Carlo) \cite{duane1987hybrid} and the No U-Turn Sampler (NUTS) \cite{hoffman2014no} now lie at the heart of MCMC software packages,  including Stan \cite{carpenter2017stan} and PyMC3 \cite{salvatier2016probabilistic}. \\

MCMC, however, does have certain disadvantages. By fundamentally relying on the convergence of a single Markov chain to its stationary distribution, MCMC approaches are poorly suited to parallel processing\footnote{The authors note that some variants of MCMC do attempt to exploit parallel processing - see TMCMC \cite{ching2007transitional} for example. However, to the best of our knowledge, none of these are able to fully exploit the potential provided by modern computing architectures.}. Moreover, samples generated before convergence of the Markov chain(s) cannot be treated as samples from the distribution of interest (herein referred to as the `target distribution') and, as such, must be discarded (a process typically referred to as `burn-in' or `convergence to the typical set'). The current paper considers, instead, an alternative to MCMC known as Sequential Monte Carlo (SMC) samplers. \\

SMC samplers\footnote{Not to be confused with SMC \emph{methods}, which  we define to be  a collection of approaches that include, for example, the particle filter \cite{arulampalam2002tutorial}.} were first proposed in \cite{del2006sequential}. They can fulfil the same role as MCMC in that, over successive iterations, they can be used to realise Monte-Carlo estimates of statistical moments associated with an arbitrary probability distribution. SMC samplers have been used to aid rare event estimation \cite{cerou2012sequential}, approximate Bayesian (a.k.a `likelihood free') computation \cite{filippi2013optimality, del2012adaptive, peters2012sequential, bonassi2015sequential, toni2009approximate}, inference from growing sets of data \cite{chopin2002sequential, green2017estimating} and Bayesian model selection \cite{zhou2016toward, jasra2008interacting} while the papers  \cite{kantas2014sequential, lee2010utility} have explored the applications of SMC samplers on parallel computing architectures. \\

SMC samplers differ from MCMC in that they are fundamentally based on an importance sampling approach, whereby a population of weighted samples evolve over a number of iterations. SMC samplers have additional `settings', not available to MCMC, that can, potentially, be used to improve the performance of the resulting sampler. One of these settings is the so-called L-kernel - a user defined conditional probability density function that appears in the general SMC implementation \cite{del2006sequential}. \\

The current paper  explains  the optimal choice for this L-kernel.  The novelty of the paper lies in the description of approximation schemes for implementing approximately optimal L-kernels for cases where the target distribution is uni-modal or multi-modal.  \\

The paper is organised as follows. Section \ref{sec:SMC_samplers} introduces SMC samplers and the role of the L-kernel before, in Section \ref{sec:optL}, the optimal L-kernel is defined. Methods for implementing an approximately optimal L-kernel are then described in Section \ref{sec:approx_opt} before the approach is illustrated on a number of case studies in Section \ref{sec:case_studies}. Future work and main conclusions are then described in Sections \ref{sec:discussion} and \ref{sec:conclusions} respectively. \\

\section{SMC samplers}\label{sec:SMC_samplers}

\subsection{General algorithm}

This section provides an introduction to SMC samplers and establishes the role of the L-kernel in a general SMC sampler implementation. \\

Ultimately, the goal of the algorithm described in the current section is to  estimate the expectations of functions with respect to a target distribution,  $\pi(\x)$, where $\x \in \mathbb{R}^D$.  We note that the approach is applicable to scenarios where the target distribution is only known up to a constant of proportionality (which is often the case in Bayesian inference problems\footnote{For clarity, generating samples from the distribution $p(\x, \y) = p(\y | \x) p(\x)$ for fixed $\y$ cannot be approached in the same manner as generating samples from $p(\x | \y)$, as $p(\x | \y)$ is proportional, not equal, to $p(\y | \x) p(\x)$.}). Consequently, in the following, we use $\pi^*(\x)$ to represent an unnormalised target distribution, such that

\begin{equation}
    \pi(\x) = \frac{\pi^*(\x)}{\int \pi^*(\x) d\x}
\end{equation}
The first iteration of an SMC sampler is identical to a standard importance sampling approach. Specifically, $N$ samples of $\x$ are generated from an initial proposal distribution, $q(\x)$, such that

\begin{equation}
    \{ \X_1^1, ..., \X_1^N \} \sim q(\x)
\end{equation}
 For the case where the normalising constant of the target distribution is unknown, the importance weight associated with each sample is

\begin{equation}
    w^*(\X_1^i) = \frac{\pi^*(\X_1^i)}{q(\X_1^i)}
\end{equation}
These weighted samples can be used to estimate statistical quantities relating to the target distribution, as

\begin{equation}
    \int f(\x) \pi(\x) d\x = \frac{\int f(\x) \pi^*(\x) d\x}{\int \pi^*(\x) d\x} \approx \frac{\sum_{i=1}^N w^*(\X_1^i)f(\X_1^i)}{\sum_{j=1}^N w^*(\X_1^j)}
\end{equation}
(where $f(\x)$ is the quantity of interest). \\

As with any importance sampling scheme, the so-called `effective sample size' \cite{kong1992note} can be used to analyse sample coverage (such that, for example, a low effective sample size implies that only a small number of samples are associated with relatively large importance weights). If the effective sample size is found to be below a pre-defined threshold, a procedure known as \emph{resampling} is undertaken. Resampling involves generating a new set of samples by sampling, with replacement, from the current set of samples. The probability that a sample from the current set, $\X_1^i$,  will become part of the new set is proportional to that sample's importance weight ($w^*(\X_1^i)$). Consequently, resampling tends to produce replicas of samples that are  sparsely separated relative to the  probability density of $\pi(\x)$,  i.e., where the density is high,  while also removing samples that are in  regions of low  probability density of $\pi(\x)$.  In all of the following, resampling is employed if the ratio of effective sample size to the number of samplers per iteration falls below 0.5.  \\

Up until this point, the SMC sampler has essentially acted as a standard importance sampler. In the next iteration, however, a new set of samples are proposed conditional on the current set. Specifically:

\begin{equation}
    \X_2^i \sim \underrightarrow{q}(\x_2 | \X_1^i), \qquad i = 1,...,N
\end{equation}
 where, here, $\underrightarrow{q}(\cdot | \cdot)$ is referred to as the `general proposal distribution'.  The main advantage of this iterative approach is that, if resampling has occurred in the previous iteration, the algorithm allows new samples to be generated in regions of interest (i.e., regions where samples from the previous iteration were associated with relatively high importance weights).  The importance weights associated with the SMC sampler's second iteration are

\begin{equation}
    w^*(\X_1^i, \X_2^i) = \frac{\pi^*_{1:2}(\X_1^i, \X_2^i)}{\underrightarrow{q}(\X_2^i | \X_1^i) q(\X_1^i)}
\end{equation}
At this point it is important to note that the weighted samples $\{ \X_1^i, \X_2^i \}_{i=1}^N$ now target $\pi_{1:2}(\x_1, \x_2)$. Practically speaking, it is only necessary for the most recent set of samples, $\{ \X_2^i \}_{i=1}^N$, to target $\pi(\x)$. Consequently, the user has significant flexibility in the choice of $\pi^*_{1:2}(\x_1, \x_2)$ subject to the condition that, when normalised, $\pi_{1:2}(\x_1, \x_2)$ is a valid probability density function (PDF) where

\begin{equation}
    \int \pi^*_{1:2}(\x_1, \x_2) d\x_1 \propto \pi(\x_2)
\end{equation}
As described in \cite{del2006sequential}, the user may choose to set

\begin{equation}
    \pi^*_{1:2}(\x_1, \x_2) = \pi^*(\x_2)L_1(\x_1 | \x_2)
    \label{eq:target01}
\end{equation}
where $L_1(\x_1 | \x_2)$, the \emph{L-kernel}, is any valid PDF. The choice of L-kernel gives a degree of flexibility that, potentially, can be used to significantly influence the performance of the SMC sampler.\\

From equation (\ref{eq:target01}) it is now possible to write the weights of each sample, $\{ \X_1^i, \X_2^i \}$, as

\begin{equation}
    w^*(\X_1^i, \X_2^i) = \frac{\pi^*(\X_2^i) L_1(\X_1^i | \X_2^i)}{\underrightarrow{q}(\X_2^i | \X_1^i) q(\X_1^i)} = \frac{\pi^*(\X_2^i)}{\pi^*(\X_1^i)} \frac{L_1(\X_1^i|\X_2^i)}{\underrightarrow{q}(\X_2^i|\X_1^i)} w^*(\X_1^i)
\end{equation}
In the general case then, when the SMC sampler has run for $k$ iterations, the distribution

\begin{equation}
    \pi^*_{1:k}(\x_{1:k}) = \pi^*(\x_k) \prod_{k'=2}^{k} L_{k'-1}(\x_{k'-1} | \x_{k'})
\end{equation}
is being targeted by the weighted samples $\{ \X_{1:k}^i \}_{i=1}^N$ whose importance weights can be found iteratively from

\begin{equation}
    w^*(\X_{1:k}^i) = \frac{\pi^*(\X_k^i)}{\pi^*(\X_{k-1}^i)} \frac{L_{k-1}(\X_{k-1}^i | \X_k^i)}{\underrightarrow{q}(\X_k^i | \X_{k-1}^i)}
    w^*(\X_{1:k-1}^i)
    \label{eq:SMC_general_weights}
\end{equation}
For notational purposes we note that, if the normalising constant of the target distribution is known, then the importance weights associated with the first and $k$th iteration of the SMC sampler would be calculated according to

\begin{equation}
    w(\X_1^i) = \frac{\pi(\X_1^i)}{q(\X_1^i)}
\end{equation}
and

\begin{equation}
    w(\X_{1:k}^i) = \frac{\pi(\X_k^i)}{\pi(\X_{k-1}^i)} \frac{L_{k-1}(\X_{k-1}^i | \X_k^i)}{\underrightarrow{q}(\X_k^i | \X_{k-1}^i)}
    w(\X_{1:k-1}^i)
\end{equation}
respectively.

\subsection{Recycling}\label{sec:recycling}
At the $k$th iteration of an SMC sampler, the quantity of interest is estimated according to

\begin{equation}
    \hat{f}_k = \frac{\sum_{i=1}^N w^*(\X_{1:k}^i) f(\X_k^i)}{\sum_{j=1}^N w^*(\X_{1:k}^j)}
\end{equation}
Recycling schemes involve combining the estimates $\hat{f}_1, ..., \hat{f}_k$ to realise an overall estimate, denoted $\hat{f}$. In the following, $\hat{f}$ is realised using a linear combination of $\hat{f}_1,...,\hat{f}_k$ such that

\begin{equation}
    \hat{f}
    = \sum_{k'=1}^k c_{k'} \hat{f}_{k'}
    = \sum_{k'=1}^k \sum_{i=1}^N \left( c_{k'} \frac{w^*(\X_{1:k'}^i)}{\sum_{j=1}^N w^*(\X_{1:k'}^j)} \right) f(\X_{k'}^i)
\end{equation}
where

\begin{equation}
    \sum_{k'=1}^k c_{k'} = 1 \quad \text{and} \qquad 0 \leq c_{k'} \leq 1, \quad k'=1,...,k
    \label{eq:lambda_constraints}
\end{equation}
In the following, the optimal choice of the constants $c_1,...,c_k$ are defined as those that maximise the effective sample size:

\begin{equation}
    \left[
        \sum_{k'=1}^k \sum_{i=1}^N \left(
            c_{k'} \frac{w^*(\X_{1:k'}^i)}{ \sum_{j=1}^N w^*(\X_{1:k'}^j)}
        \right)^2
    \right]^{-1}
\end{equation}
subject to the constraints described by equation (\ref{eq:lambda_constraints}). As described in \cite{nguyen2015efficient}, this leads to

\begin{equation}
    c_k^{\text{opt}} = \frac{l_k}{\sum_{k'=1}^k l_{k'}} \quad \text{where} \quad l_k = \frac{( \sum_{j=1}^N w^*(\X_{1:k}^j) )^2}{\sum_{i=1}^N w^*(\X_{1:k}^i)^2}
\end{equation}
where $c_k^{\text{opt}}$ represents the optimal choice of $c_k$.

\section{Optimal L-kernel}\label{sec:optL}
The L-kernel can be set such that

\begin{equation}
    L_{k-1}(\x_{k-1} | \x_k)= \underrightarrow{q}(\x_{k-1} | \x_{k}), \qquad \forall \x_{k-1}, \x_k
    \label{eq:L_equals_q}
\end{equation}
(this strategy, referred to here as the `forward proposal L-kernel', was undertaken in \cite{green2017estimating}). Using a forward proposal L-kernel, the importance weights of the SMC algorithm are updated according to 

\begin{equation}
    w(\X_{1:k}^i) = \frac{\pi(\X_k^i)}{\pi(\X_{k-1}^i)} \frac{\underrightarrow{q}(\X_{k-1}^i | \X_{k}^i)}{\underrightarrow{q}(\X_k^i | \X_{k-1}^i)}
    w(\X_{1:k-1}^i)
    \label{eq:SMC_MH}
\end{equation}
which bares similarities to the Metropolis-Hastings acceptance rule\footnote{For clarity, in the notation of the current paper, the Metropolis-Hastings algorithm accepts $\x_k$ as the new state of the Markov chain with probability $\min \left\{ 1, \frac{\pi(\x_k)}{\pi(\x_{k-1})} \frac{\underrightarrow{q}(\x_{k-1} | \x_{k})}{\underrightarrow{q}(\x_k | \x_{k-1})} \right\}$.}. While use of the forward proposal L-kernel appears mathematically convenient we stress that it is, in fact, generally suboptimal for the sampler itself. \\

The current section describes a mathematical definition of the \emph{optimal} L-kernel. Specifically, in Section \ref{sec:optL_simplified}, we begin by discussing the optimal L-kernel for a simplified SMC sampler that runs over 2 iterations only. This result is then demonstrated on a simple illustrative example in Section \ref{sec:optL_example} before a more general derivation of the optimal L-kernel is outlined in Section \ref{sec:optL_general}.  The authors note that the optimal L-kernel was originally shown in \cite{del2006sequential} and that the novelty of the current paper lies in the implementation strategy described in Section \ref{sec:approx_opt}. \\

\subsection{Simplified case}\label{sec:optL_simplified}
Here we restrict our discussion to the case where the normalising constant of the target distribution is known and consider an SMC sampler that is being used to estimate $\int f(\x) \pi(\x) d\x$ (discussion of the case where the normalising constant is unknown is postponed until the end of Section \ref{sec:optL_general}). Furthermore, for the sake of simplicity, we initially consider a sampler that is run over 2 iterations only and where no resampling takes place (resampling is also addressed in Section \ref{sec:optL_general}). \\

At iteration 2, the sampler's estimate of $\int f(\x) \pi(\x) d\x$ is

\begin{equation}
	\tilde{f}_2 = \frac{1}{N} \sum_{i=1}^N \frac{\pi(\X_2^i)L_1(\X^i_1|\X^i_2)}{\underrightarrow{q}(\X_2^i | \X_1^i)q(\X_1^i)}f(\X_2^i), \qquad \X_{1:2}^i \sim \underrightarrow{q}(\x_2|\x_1)q(\x_1)
    \label{eq:2_iteration_estimate}
\end{equation}
The variance of the estimate shown in equation (\ref{eq:2_iteration_estimate}) is

\begin{equation}
	\Var[\tilde{f}_2] = \E[\tilde{f}_2^2] + \text{const}
\end{equation}
where

\begin{equation}
	\E[\tilde{f}_2^2] = \frac{1}{N} \sum_{i=1}^N \int \left( \frac{\pi(\X_2^i)L_1(\X_1^i|\X_2^i)}{\underrightarrow{q}(\X_2^i|\X_1^i)q(\X_1^i)} f(\X_2^i) \right)^2 \underrightarrow{q}(\X_2^i|\X_1^i)q(\X_1^i) d\X_{1:2}^i
\end{equation}

\begin{equation}
	= \int \frac{\pi(\x_2)^2 L_1(\x_1|\x_2)^2}{\underrightarrow{q}(\x_2|\x_1)q(\x_1)} f(\x_2)^2 d\x_{1:2}
    \label{eq:2_iteration_sample_variance}
\end{equation}
We now aim to find the L-kernel, $L_1(\x_1|\x_2)$, that minimises the variance of the estimator (i.e., minimises equation (\ref{eq:2_iteration_sample_variance})) subject to the constraint that

\begin{equation}
    \int L_1(\x_1|\x_2) d\x_1 = 1 \qquad \forall \x_2
    \label{eq:2_iteration_constraint}
\end{equation}
(therefore ensuring that the L-kernel is a valid PDF). This constrained optimisation problem can be addressed using a Lagrange multiplier approach were, to ensure that the constraint in equation (\ref{eq:2_iteration_constraint}) is enforced, the following Lagrangian function is defined:

\begin{equation}
	\E[\tilde{f}_2^2] - \int \lambda(\x_2') \left( \int L_1(\x_1|\x_2') d\x_1 - 1 \right) d\x_2'
\end{equation}
where the $\lambda$s are Lagrange multipliers. The optimal L-kernel therefore satisfies the following:

\begin{equation}
	\frac{\partial}{\partial L_1} \left[
		\E[\tilde{f}_2^2] - \int \lambda(\x_2') \left( \int L_1(\x_1|\x_2') d\x_1 - 1 \right) d\x_2'
	\right]=0
    \label{eq:2_iteration_derivative}
\end{equation}
(where $\frac{\partial}{\partial L_1}$ is a functional derivative). Writing the optimal L-kernel as $L_1^{\text{opt}}  (\x_1|\x_2)$ then, from equation (\ref{eq:2_iteration_derivative}), it follows that

\begin{equation}
	2 \int \frac{\pi(\x_2)^2  L_1^{\text{opt}}  (\x_1|\x_2)}{\underrightarrow{q}(\x_2|\x_1)q(\x_1)} f(\x_2)^2 d\x_{1:2} + \int \lambda(\x_2') d\x_2' = 0
\end{equation}
which is satisfied when

\begin{equation}
	2 \frac{\pi(\x_2)^2  L_1^{\text{opt}}  (\x_1|\x_2)}{\underrightarrow{q}(\x_2|\x_1)q(\x_1)} f(\x_2)^2 + \lambda(\x_2) = 0
\end{equation}
therefore, noting that $\lambda(\x_2)$ is constant for fixed $\x_2$, it follows that

\begin{equation}
	 L_1^{\text{opt}}  (\x_1|\x_2) \propto \frac{\underrightarrow{q}(\x_2|\x_1)q(\x_1)}{f(\x_2)^2 \pi(\x_2)^2} \propto \underrightarrow{q}(\x_2|\x_1)q(\x_1)
    \label{eq:2_optL}
\end{equation}
Equation (\ref{eq:2_optL}) shows the optimal L-kernel (i.e., the L-kernel that would minimise the sample variance of the estimator) for an SMC sampler that has been applied over 2 iterations. To realise an analytical expression for $ L_1^{\text{opt}}  (\x_1|\x_2)$, one must rearrange $\underrightarrow{q}(\x_2|\x_1)q(\x_1)$ into the form $\underleftarrow{q}(\x_1 | \x_2)$  where $\underleftarrow{q}(\x_1 | \x_2) d\x_1$ describes the probability that, given $\x_2$, there had been a sample in the interval $[\x_1, \x_1+d\x_1]$ at the first iteration of the SMC sampler.  To emphasise this point, an illustrative example is given in the following section. \\

\subsection{Illustrative Example}\label{sec:optL_example}
Here we consider the case where $x \in \mathbb{R}$ and where the aim is to estimate the mean of the distribution $\pi(x) = \mathcal{N}(x; 1, 1)$. The SMC sampler is run for 2 iterations, whereby samples are proposed according to

\begin{equation}
	q(x_1) = \mathcal{N}(x_1; 0, 1), \qquad \underrightarrow{q}(x_2|x_1) = \mathcal{N}(x_2; x_1, 1)
\end{equation}
In this case, using standard expressions for Gaussian distributions, a closed-form expression for $\underleftarrow{q}(x_1|x_2)$ can be shown to be

\begin{equation}
	\underleftarrow{q}(x_1|x_2) = \mathcal{N}\left( x_1; \frac{x_2}{2}, \frac{1}{2} \right)
\end{equation}
Note that, in this notation, $\underrightarrow{q}(x_2|x_1)dx_2$ represents the probability of proposing a sample in the region $[x_2, x_2 + dx_2]$ given that the algorithm's state is currently $x_1$. Conversely, if the algorithm's state is currently $x_2$, $\underleftarrow{q}(x_1|x_2)dx_1$ represents the probability that the algorithm's previous state was in the region $[x_1, x_1 + dx_1]$. Recalling equation (\ref{eq:2_optL}), the optimal L-kernel for this simple case study is therefore

\begin{equation}
     L_1^{\text{opt}}  (x_1|x_2) = \mathcal{N}\left( x_1; \frac{x_2}{2}, \frac{1}{2} \right)
    \label{eq:simple_example_optL}
\end{equation}

\begin{figure}[H]
	\centering
	\includegraphics[scale=0.9, trim={0 1cm 0 0},clip]{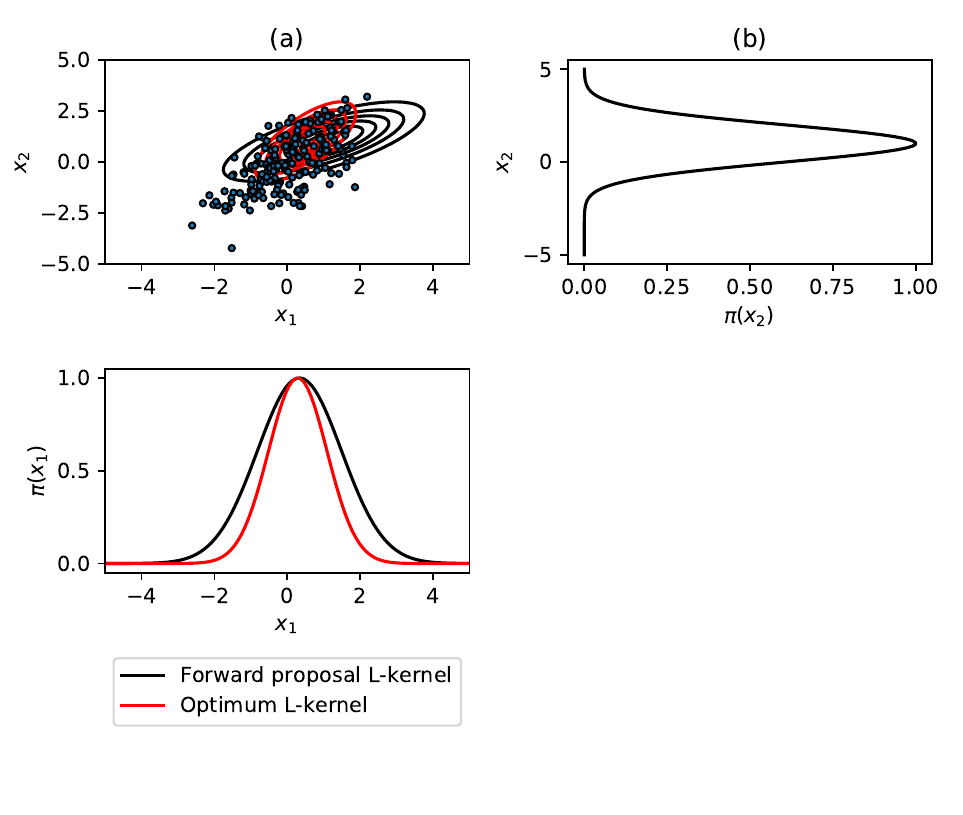}
    \caption{Illustrative example described in Section \ref{sec:optL_example}. The contours in (a) and (c) represent the target distribution, $\pi_{1:2}(\x_{1:2})$, for the case where  $L_1(x_1|x_2)=\mathcal{N}(x_1; x_2, 1)$ and $L_1(x_1|x_2)=\mathcal{N}\left(x_1; \frac{x_2}{2}, \frac{1}{2} \right)$ respectively. }
	\label{fig:simple_optL}
\end{figure}

The filled circles in  Figure \ref{fig:simple_optL} (a) show 200 samples that were generated by an SMC sampler over two iterations. The contours shown in Figure \ref{fig:simple_optL} (a) illustrate the multi-dimensional target distribution, $\pi_{1:2}(\x_{1:2})$, for the cases where the forward proposal L-kernel is used ($L_1(x_1 | x_2) = \underrightarrow{q}(x_1 | x_2)$) and where the optimal L-kernel (equation (\ref{eq:simple_example_optL})) is used. Figure \ref{fig:simple_optL} (b) shows the marginal of the two contours over $x_2$. We note that, as a result of the constraint outlined by equation (\ref{eq:2_iteration_constraint}), the marginal of the  contours  in Figure \ref{fig:simple_optL} (a) are both equal to $\pi(x_2)$ (i.e., $\int \pi_{1:2}(\x_{1:2}) dx_1 = \pi(x_2)$.) Figure \ref{fig:simple_optL} (c), however, shows that the marginal of the contours over $x_1$ are not necessarily the same.\\

Figure \ref{fig:simple_optL} helps to give an intuitive impression of how the optimal L-kernel improves sampler performance. The optimal-L contour in Figure \ref{fig:simple_optL}  (a)  appears to be positioned such that it encloses a relatively large number of samples, thus maintaining a high effective sample size whilst still having the property that the marginal of $\pi_{1:2}(\x_{1:2})$, over $x_1$, is equal to $\pi(x_2)$. In this way, it is possible to envisage how using the optimal L-kernel can help to prevent relatively low values of effective sample size from occurring. \\

In the following section, a generalised expression for the optimal L-kernel is derived and a more mathematically rigorous explanation as to its performance enhancements is outlined. \\

\subsection{General Case}\label{sec:optL_general}

Consider now, a general SMC sampler that has run over $k$ iterations and where $\x \in \mathbb{R}^D$. If no resampling has taken place before iteration $k$ then the estimate, $\tilde{f}_k$, is realised according to

\begin{equation}
    \tilde{f}_k =
    \frac{1}{N} \sum_{i=1}^N \frac{\pi(\X_k^i) \prod_{k'=2}^k L_{k'-1}(\X_{k-1}^i | \X_k^i)}{q(\X_1^i) \prod_{k'=2}^k \underrightarrow{q}(\X_{k'}^i|\X_{k'-1}^i)} f(\X_k^i),
    \label{eq:estimate_iteration3}
\end{equation}
where

\begin{equation}
    \X_{1:k}^i \sim q(\x_1) \prod_{k'=2}^k \underrightarrow{q}(\x_{k'}|\x_{k'-1})
\end{equation}
In general, the optimal L-kernel is

\begin{equation}
	L_{k-1}^{\text{opt}} (\x_{k-1} | \x_k) = \underleftarrow{q}(\x_{k-1}| \x_k)
    \label{eq:optL}
\end{equation}
As noted in \cite{del2006sequential},  setting each of the L-kernels equal to their optimal value, the importance weights associated with a single `trajectory' ($\x_{1:k}$) of a generalised SMC sampler are therefore

\begin{equation}
	w(\x_{1:k}) = \frac{\pi_{1:k}(\x_{1:k})}{q(\x_{1:k})}
		= \frac{\pi(\x_k) \prod_{k'=2}^k L_{k'-1}^{\text{opt}}(\x_{k'-1}|\x_{k'})}{q(\x_k) \prod_{k'=2}^{k} \underleftarrow{q}(\x_{k'-1}|\x_{k'})}
        = \frac{\pi(\x_k)}{q(\x_k)}
    \label{eq:lower_dimensional_space}
\end{equation}
Equation (\ref{eq:lower_dimensional_space}) highlights how, by using the optimal L-kernel, the samples can now be viewed as inhabiting a lower dimensional space. Specifically, the importance weights associated with each trajectory ($\x_{1:k}$) can now be calculated based purely with regards to the proposal distribution $q(\x_k) = \int q(\x_{1:k}) d\x_{1:k-1}$. \\

If resampling last occurred at iteration $l < k$ then the samples $\X_l^1, ..., \X_l^N$ can be considered as (approximate) samples from $\pi(\x_l)$. As a result, if we consider samples only after iteration $l$, we have

\begin{equation}
    \tilde{f}_k \approx
    \frac{1}{N} \sum_{i=1}^N \frac{\pi(\X_k^i) \prod_{k'=l+1}^k L_{k'-1}(\X_{k-1}^i | \X_k^i)}{\pi(\X_l^i) \prod_{k'=l+1}^k \underrightarrow{q}(\X_{k'}^i|\X_{k'-1}^i)} f(\X_k^i), 
\end{equation}
where, now, we have 

\begin{equation}
    q(\x_k) = \int \pi(\x_l) \prod_{k'=l+1}^k \underrightarrow{q}(\x_{k'}|\x_{k'-1}) d\x_{l:k-1}
\end{equation}
The optimum L-kernel, however, is still as described by equation (\ref{eq:optL})\footnote{The authors note that the notation used in the current paper doesn't necessarily capture the fact that the samples $\X_l^1,...\X_l^N$ are approximate samples from the target, $\pi$. The aim, at this point of the manuscript, is to illustrate that resampling does not alter the novel optimum L-kernel approximation schemes that are proposed in Section \ref{sec:approx_opt}. For a more complete notation, the paper \cite{del2006sequential} is recommended.}. \\

Finally, for the case where the normalising constant of the target distribution is unknown, at iteration $k$, estimates of the quantity of interest are realised from

\begin{equation}
    \tilde{f}_k = \frac{\sum_{i=1}^N w^*(\X_{1:k}^i) f(\X_k^i)}{\sum_{j=1}^N w^*(\X_{1:k}^j)}
    \label{eq:optL_unnorm_target}
\end{equation}
In this case, the optimal L-kernel is still equal to that shown in equation (\ref{eq:optL}) as this approach will minimise the variance of the estimates of both the numerator and the denominator of equation (\ref{eq:optL_unnorm_target}).

\section{Estimating the optimal  L-kernel}\label{sec:approx_opt}
General closed-form expressions for the optimal L-kernel are usually intractable. As a result, in the current section, we present an approach that allows the implementation of an \emph{approximately} optimal L-kernel. More specifically, Section \ref{sec:gaussian_approx} describes an approximation that is suitable for situations where $q(\x_{k-1}, \x_k)$ can be approximated as a Gaussian distribution while Section \ref{sec:gaussian_mixture_approx} describes a more general approximation that can be applied when, for example, the target distribution has multiple modes. \\

\subsection{Gaussian Approximation}\label{sec:gaussian_approx}
If it assumed that

\begin{equation}
	q(\x_{k-1}, \x_k) \approx  \hat{q}(\x_{k-1}, \x_k) =  \mathcal{N}\left(
		\left(
		\begin{array}{c}
			\x_{k-1} \\
			\x_k \\
		\end{array}
		\right);
		\left(
		\begin{array}{c}
			\bmu_{k-1} \\
			\bmu_k \\
		\end{array}
		\right),
		\left[
		\begin{array}{cc}
			\bSigma_{k-1,k-1} & \bSigma_{k-1,k} \\
			\bSigma_{k,k-1} & \bSigma_{k,k} \\
		\end{array}
		\right]
	\right)
    \label{eq:q_Gaussian_approximation}
\end{equation}
then, using established properties of Gaussian distributions, it is possible to show that

\begin{equation}
	 \hat{\underleftarrow{q}}(\x_{k-1}| \x_k) =  \mathcal{N}\left(
		\x_{k-1}; \bmu_{k-1|k}, \bSigma_{k-1|k}
	\right)
    \label{eq:optL_Gaussian_approximation}
\end{equation}
where

\begin{equation}
	\bmu_{k-1|k} = \bmu_{k-1} + \bSigma_{k-1,k} \bSigma^{-1}_{k,k} (\x_k - \bmu_k)
\end{equation}

\begin{equation}
	\bSigma_{k-1|k} = \bSigma_{k-1,k-1} - \bSigma_{k-1,k} \bSigma^{-1}_{k,k} \bSigma_{k,k-1}
\end{equation}
Consequently, by using existing samples of $(\x_{k-1},\x_k$) to estimate the parameters of the Gaussian approximation described in equation (\ref{eq:q_Gaussian_approximation}), an approximately optimal L-kernel can be realised by setting $L_{k-1}(\x_{k-1}|\x_k)$ equal to  $\hat{\underleftarrow{q}}(\x_{k-1}|\x_k)$  (where the latter is defined in equation (\ref{eq:optL_Gaussian_approximation})). \\

\subsection{Mixture-of-Gaussian Approximation}\label{sec:gaussian_mixture_approx}
For situations where a Gaussian approximation of $q(\x_{k-1},\x_k)$ is poor, we can instead attempt to use a Mixture-of-Gaussian approximation. Specifically, this approach involves using samples of $(\x_{k-1},\x_k)$ to realise the approximation

\begin{equation}
	 \hat{q}(\x_{k-1}, \x_k) =
    \sum_m \Pr(m)
    \mathcal{N}\left(
		\left(
		\begin{array}{c}
			\x_{k-1} \\
			\x_k \\
		\end{array}
		\right);
		\left(
		\begin{array}{c}
			\bmu_{k-1}^{(m)} \\
			\bmu_k^{(m)} \\
		\end{array}
		\right),
		\left[
		\begin{array}{cc}
			\bSigma_{k-1,k-1}^{(m)} & \bSigma_{k-1,k}^{(m)} \\
			\bSigma_{k,k-1}^{(m)} & \bSigma_{k,k}^{(m)} \\
		\end{array}
		\right]
	\right)
\end{equation}
where $m$ indexes each mixture component. Such an approach has the advantage that, once a Gaussian Mixture Model has been fitted to the samples $(\x_{k-1},\x_k)$, closed-form expressions for $\hat{\underleftarrow{q}}(\x_{k-1}|\x_k)$ can be realised. Specifically, it can be shown that

\begin{equation}
	 \hat{\underleftarrow{q}}(\x_{k-1}| \x_k) =  \sum_m \Pr(m| \x_k) \mathcal{N}\left(
		\x_{k-1}; \bmu_{k-1|k}^{(m)}, \bSigma_{k-1|k}^{(m)}
	\right)
\end{equation}
where

\begin{equation}
	\bmu_{k-1|k}^{(m)} = \bmu_{k-1}^{(m)} + \bSigma_{k-1,k-1}^{(m)} \left(\bSigma^{(m)}_{k,k}\right)^{-1} (\x_k - \bmu_k^{(m)})
\end{equation}

\begin{equation}
	\bSigma_{k-1|k}^{(m)} = \bSigma_{k-1,k-1}^{(m)} - \bSigma_{k-1,k}^{(m)} \left(\bSigma^{(m)}_{k,k}\right)^{-1} \bSigma^{(m)}_{k,k-1}
\end{equation}
and

\begin{equation*}
    \Pr(m | \x_k) = \frac{p(\x_k | m) \Pr(m)}{\sum_{m'} p(\x_k | m') \Pr(m')}
\end{equation*}

\begin{equation}
    = \frac{ \mathcal{N}(\x_k | \bmu_k^{(m)}, \bSigma_{k,k}^{(m)}) \Pr(m)}{\sum_{m'} \mathcal{N}(\x_k | \bmu_k^{(m')}, \bSigma_{k,k}^{(m')}) \Pr(m')}
\end{equation}
In the following examples, when employed, all Gaussian Mixture Models are trained using the Expectation Maximisation algorithm (implemented in Python's `scikit-learn' package \cite{scikit-learn}). \\

\section{Case Studies}\label{sec:case_studies}
This section details 3 case studies that demonstrate the proposed approach. Section \ref{sec:2D_toy_problem} describes a problem with a 2-dimensional target distribution, before, in Section \ref{sec:bimodal_toy_problem}, we study a scenario with a 1-dimensional \emph{bi-modal} target distribution. Finally, in Section \ref{sec:heli}, we apply the approach to a real case study where the aim is to analyse the hyperparameter uncertainty in a Gaussian Process rotorcraft simulation model. \\

\subsection{2D Toy Problem}\label{sec:2D_toy_problem}
To begin with, we study a 2D problem with target distribution

\begin{equation}
	\pi(\x) = \mathcal{N}(\x; \bmu, \bSigma)
\end{equation}
where

\begin{equation}
	\bmu = \left[
		\begin{array}{cc}
			3 & 2 \\
		\end{array}
		\right]^T
	\qquad
	\bSigma = \left[
		\begin{array}{cc}
			1 & 0 \\
			0 & 1 \\
		\end{array}
	\right]
\end{equation}
The initial proposal distribution was chosen to be $q(\x_1) = \mathcal{N}(\x_1; \boldsymbol{0}, \boldsymbol{I})$ while the general proposal distribution was chosen to be $\underrightarrow{q}(\x_k|\x_{k-1}) = \mathcal{N}(\x_k; \x_{k-1}, \boldsymbol{I})$. As mentioned in Section \ref{sec:optL}, it is often the case that the L-kernel is chosen (sub-optimally) such that

\begin{equation}
    L_{k-1}(\x_{k-1}| \x_k) = \underrightarrow{q}(\x_{k-1} | \x_k)
    \label{eq:q_L_1}
\end{equation}
(in the current example, this would be achieved by setting $L_{k-1}(\x_{k-1}|\x_k) = \mathcal{N}(\x_{k-1}; \x_k, \boldsymbol{I})$). We note again that, in all of the following case studies, the choice of L-kernel shown in equation (\ref{eq:q_L_1}) is referred to as the `forward proposal L-kernel', while the approximately optimal L-kernels that are described in Section \ref{sec:approx_opt} are referred to simply as the `optimal L-kernel'. \\

SMC samplers using both the forward proposal L-kernel and the single-Gaussian approximately optimal L-kernel were run, using  $k=100$  iterations and $N=500$ samples per iteration. The code used to obtain the following results can be run from the file `\verb|2D_toy_problem.py|' in the Github repository \url{https://github.com/plgreenLIRU/SMC_approx_optL}. \\

\begin{figure}[H]
	\centering
	\includegraphics[scale=0.9]{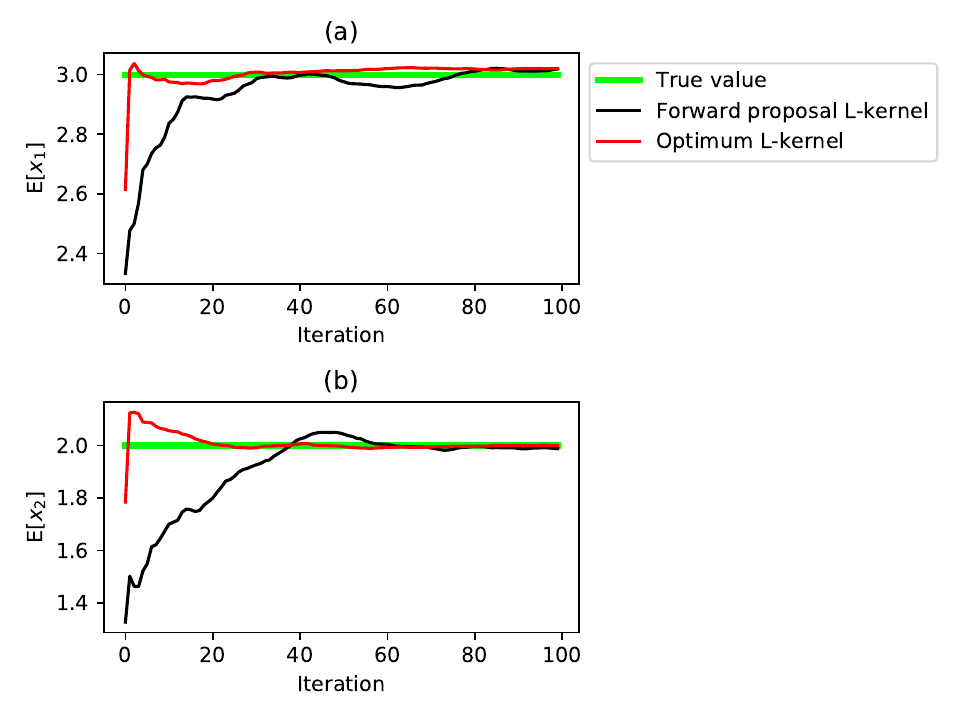}
	\caption{SMC sampler used to estimate the means for the 2D toy problem described in Section \ref{sec:2D_toy_problem}. Black lines and red lines represent results obtained using a user defined L-kernel (equation (\ref{eq:q_L_1})) and a single-Gaussian approximately optimal L-kernel respectively. Green lines show the true solution.}
	\label{fig:2D_toy_problem_mean}
\end{figure}

\begin{figure}[H]
	\centering
	\includegraphics[scale=0.9]{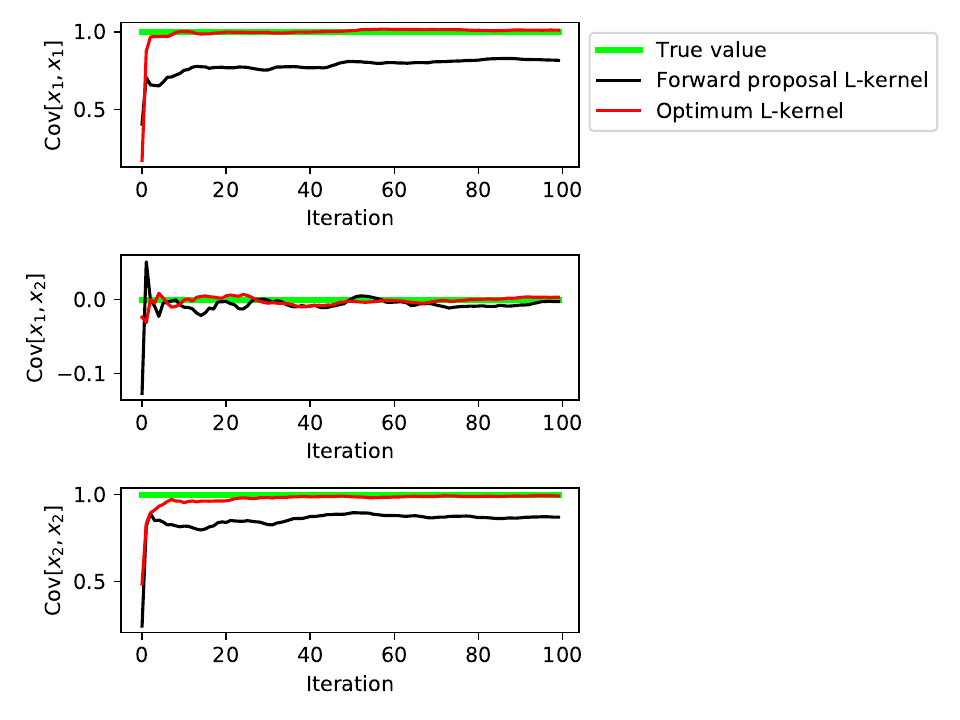}
	\caption{SMC sampler used to estimate the covariance matrix for the 2D toy problem described in Section \ref{sec:2D_toy_problem}. Black lines and red lines represent results obtained using a user defined L-kernel (equation (\ref{eq:q_L_1})) and a single-Gaussian approximately optimal L-kernel respectively. Green lines show the true solution.}
	\label{fig:2D_toy_problem_cov}
\end{figure}

\begin{figure}[H]
	\centering
	\includegraphics[scale=0.9]{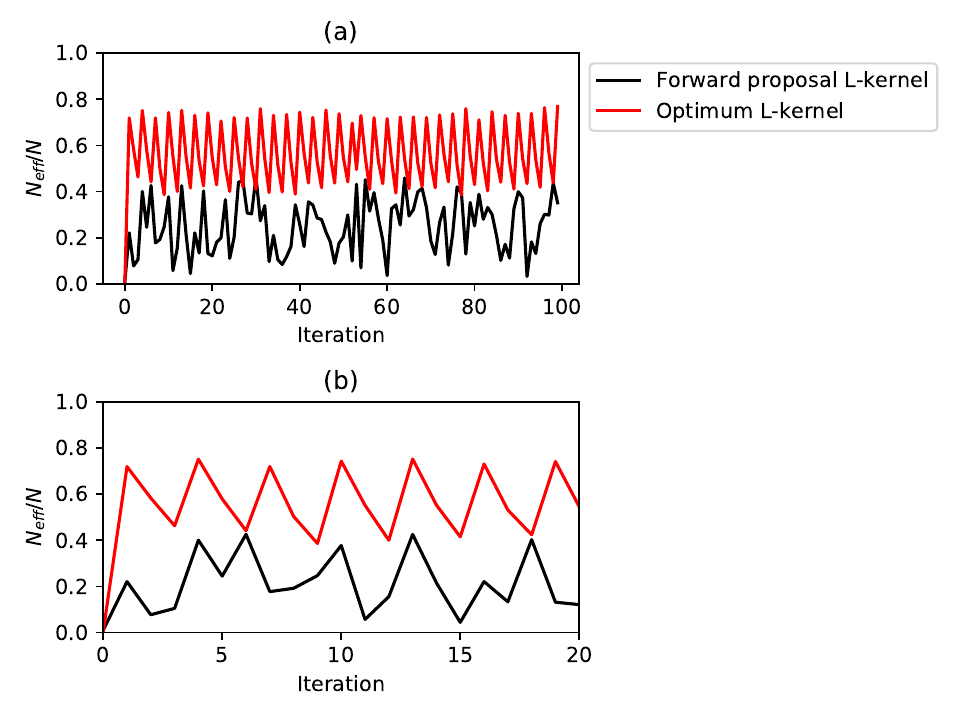}
	\caption{Effective sample size of SMC samplers applied to the 2D toy problem described in Section \ref{sec:2D_toy_problem}. Black lines and red lines represent results obtained using a user defined L-kernel (equation (\ref{eq:q_L_1})) and a single-Gaussian approximately optimal L-kernel respectively. Panel (a) shows the effective sample size over the full  100  iterations, while panel (b) shows a zoomed-in portion of the results. }
	\label{fig:2D_toy_problem_Neff}
\end{figure}

Figures \ref{fig:2D_toy_problem_mean} and \ref{fig:2D_toy_problem_cov} show the estimates of the target mean and elements of the target covariance matrix, respectively,  using the recycling scheme described in Section \ref{sec:recycling} . It is clear that, using the approximately optimal L-kernel, the  SMC algorithm has converged to the true solutions faster than the SMC implementation with the forward proposal L-kernel.  Figure \ref{fig:2D_toy_problem_Neff} shows the effective sample sizes of the two approaches. It is interesting to note that, as well as  aiding convergence , the use of an approximately optimal L-kernel has significantly reduced the number of times that resampling was required;  the forward proposal L-kernel implementation performed resampling at every iteration whilst the approximately optimal L-kernel implementation performed resampling 35 times out of the 100 iterations. Finally, Table \ref{table:2D_toy_example} shows the variance of the estimates realised using the two approaches, quantifying how, for this example, the use of an approximately optimal L-kernel reduces the variance of the algorithm's estimates. \\

\begin{table}[H]
    \centering
    \begin{tabular}{c|c|c}
         & Forward proposal L-kernel & Optimal L-kernel \\
        \hline
        $\E[x_1]$ & 0.035 & 0.012 \\
        $\E[x_2]$ & 0.091 & 0.003 \\
        $\Cov[x_1, x_1]$ & 0.014 & 0.011 \\
        $\Cov[x_1, x_2]$ & 0.013 & 0.001 \\
        $\Cov[x_2, x_2]$ & 0.021 &  0.004 \\
    \end{tabular}
    \caption{ Sample variance of the estimates realised for the example described in Section \ref{sec:2D_toy_problem}.  }
    \label{table:2D_toy_example}
\end{table}

\subsection{Bi-modal Toy Problem}\label{sec:bimodal_toy_problem}
For the second example, the aim is to target the bi-modal distribution:

\begin{equation}
    \pi(x) = \frac{1}{2} \mathcal{N}(x; \mu_1, \sigma_1^2) +
             \frac{1}{2} \mathcal{N}(x; \mu_2, \sigma_2^2)
    \label{eq:bimodal_target}
\end{equation}
where $\mu_1=-3$, $\mu_2=3$ and $\sigma_1 = \sigma_2 = 1$. The mean and variance of equation (\ref{eq:bimodal_target}) are

\begin{equation}
    \E[X] = \frac{1}{2} \sum_{c=1}^2 \mu_c = 0 \quad \text{and} \quad
    \Var[X] = \frac{1}{2} \sum_{c=1}^2 (\sigma_c^2 + \mu_c^2) = 10
\end{equation}
respectively. The initial proposal was chosen to be $q(x_1) = \mathcal{N}(x_1; 0, 3)$ while the general proposal was chosen to be $\underrightarrow{q}(x_k | x_{k-1}) = \mathcal{N}(x_k; x_{k-1}, 0.1)$. For the results shown in the remainder of this section, all SMC samplers were run for $k=1000$ iterations with $N=500$ samples per iteration. \\

Given the bi-modal nature of the target distribution, it is reasonable to expect that the joint distribution of samples $q(x_{k-1},x_k)$ will also be bi-modal. Consequently, the current example illustrates a case where we would expect a bi-modal approximation of the optimal L-kernel (implemented as described in Section \ref{sec:gaussian_mixture_approx}) to outperform the approach that approximates $q(x_{k-1},x_k)$ with a single Gaussian (Section \ref{sec:gaussian_approx}). In the following, `optimal L-kernel (1 component)'  and `optimal L-kernel (2 components)' are used to represent the uni-modal and bi-modal approximations of the optimal L-kernel respectively. As before, `forward proposal L-kernel' is used to represent the case where $L_{k-1}(x_{k-1}|x_k) = \underrightarrow{q}(x_{k-1}|x_k)$. The code used to obtain the following results can be run from the file `\verb|bimodal_toy_problem.py|' in the Github repository \url{https://github.com/plgreenLIRU/SMC_approx_optL}.\\

\begin{figure}[H]
    \centering
    \includegraphics[scale=0.9]{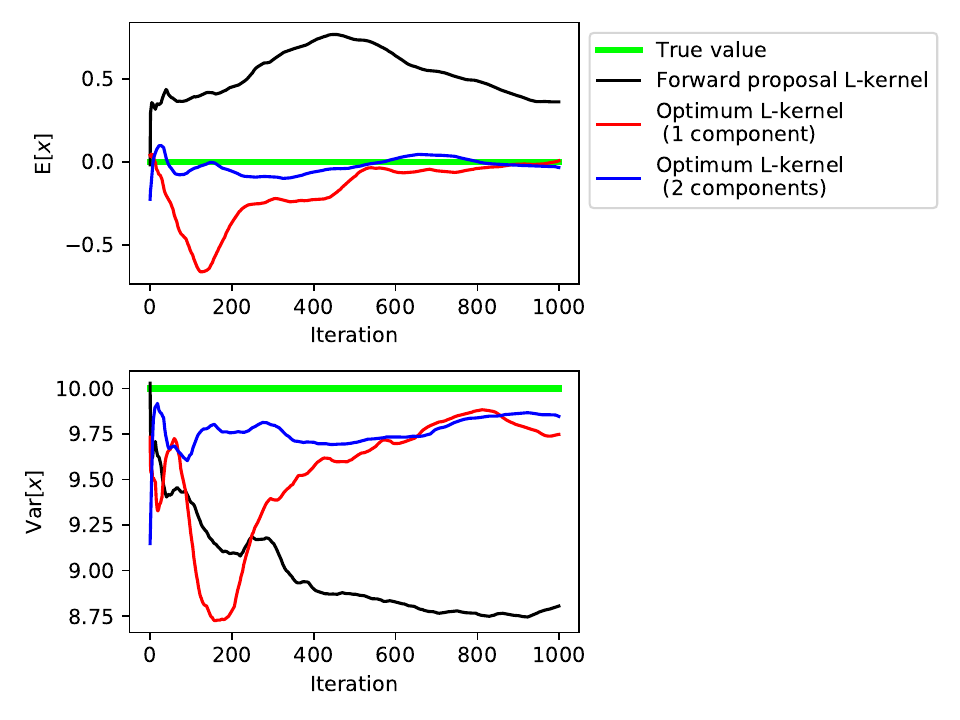}
    \caption{SMC sampler used to estimate the target mean and variance for the bi-modal toy problem described in Section \ref{sec:bimodal_toy_problem}. Black, red and blue lines represent results obtained using a user defined L-kernel, a uni-modal approximately optimal L-kernel and a bi-modal approximately optimal L-lernel respectively. Green lines show the true solution.}
    \label{fig:bimodal_toy_problem_mean_var}
\end{figure}

\begin{figure}[H]
    \centering
    \includegraphics[scale=0.9]{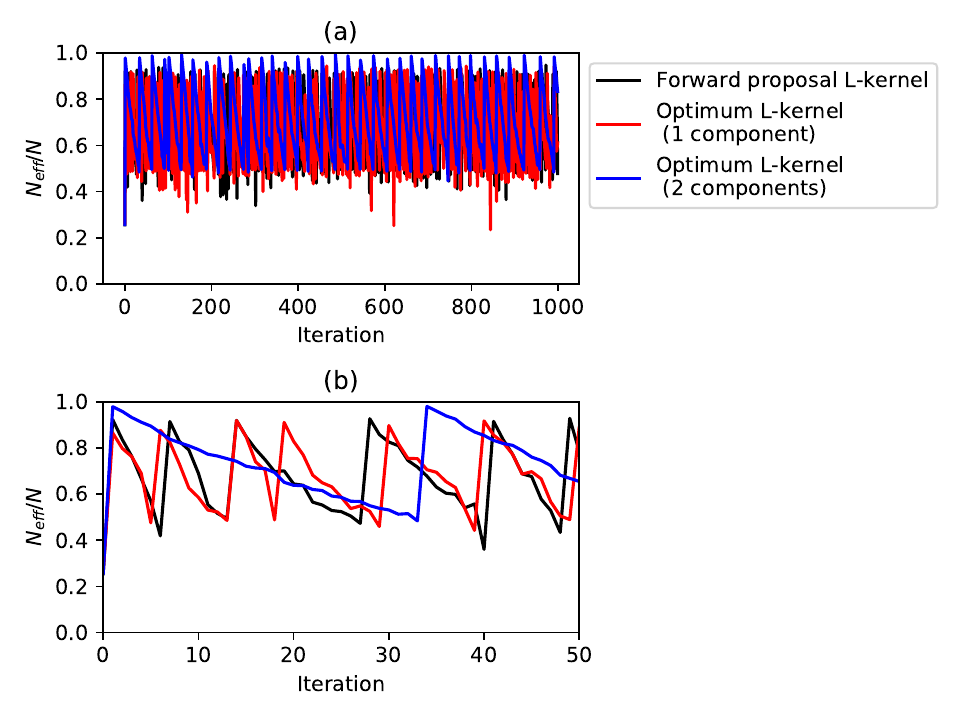}
    \caption{Effective sample size of SMC samplers applied to the bi-modal toy problem described in Section \ref{sec:bimodal_toy_problem}. Black, red and blue lines represent results obtained using a user defined L-kernel, a uni-modal approximately optimal L-kernel and a bi-modal approximately optimal L-lernel respectively. Panel (a) shows the effective sample size over the full 1000 iterations, while panel (b) shows a zoomed-in portion of the results.}
    \label{fig:bimodal_toy_problem_Neff}
\end{figure}

Figure \ref{fig:bimodal_toy_problem_mean_var} shows the resulting estimates of the target mean and variance,  using the recycling scheme described in Section \ref{sec:recycling}.  The use of a bi-modal approximately optimal L-kernel has  led to faster convergence relative to the case where the uni-modal approximately optimal L-kernel is used (although both approaches outperform the forward proposal L-kernel).  Figure \ref{fig:bimodal_toy_problem_Neff} illustrates that, by using the bi-modal approximately optimal L-kernel, the rate at which the effective sample size drops is greatly reduced;  out of the 1000 iterations the forward proposal L-kernel, uni-modal approximately optimum L-kernel and bi-modal approximately optimum L-kernel implementations performed resampling 116, 126 and 36 times respectively.  This is particularly important for the case where the target is bi-modal as, whenever resampling is required, it is possible (if improbable) that all of the samples in one particular mode of the target will be removed.  The sample variance of estimates using the 3 approaches are given in Table \ref{Table:bimodal} where it can be seen that the bi-modal approximately optimal L-kernel achieved lower sample variances than the uni-modal approximately optimal L-kernel (and both L-kernel approximation methods achieve lower sample variance than the forward proposal L-kernel). \\

\begin{table}[H]
    \centering
    \begin{tabular}{c|c|c|c}
         & Forward proposal L-kernel & \shortstack{Optimal L-kernel \\ (1 component)} & \shortstack{Optimal L-kernel \\ (2 component)} \\
        \hline
        $\E[x]$ & 0.899 & 0.635 & 0.062 \\
        $\Var[x]$ & 1.222 & 2.976 & 0.0749 \\
    \end{tabular}
    \caption{ Sample variance of the estimates realised for the example described in Section \ref{sec:bimodal_toy_problem}. }
    \label{Table:bimodal}
\end{table}

\subsection{Gaussian Process Predictions of Rotorcraft Dynamics}\label{sec:heli}

The third example considers the development of a machine-learnt flight simulator for rotorcraft. For more information about this specific example, the reader may consult \cite{jackson2019predicting}. \\

Flight simulators are a vital part of any aircraft life cycle. However, to run in real-time, simplifications to the models that underpin flight simulators often have to be made. Such simplifications can cause the behaviour of the model to differ significantly from that of the real aircraft - this is particularly true for rotorcraft where complex nonlinear phenomena can occur. Data-based approaches have the potential to provide higher fidelity flight simulation models relative to current, physical-law based approaches.  \\

The paper \cite{jackson2019predicting} details the development of a data-based flight simulator, that is trained on real flight data from a Bo105 rotorcraft\footnote{The AGARD database was delivered to the University of Liverpool as part of GARTEUR HC(AG-16) Rotorcraft Pilot Couplings \cite{Bo105_data}. Unfortunately, we do not currently have permission to publish this data alongside the current paper}. In the current work, an SMC sampler is used to help quantify uncertainties associated with parameters of this machine-learnt flight simulator model. \\

\begin{figure}[H]
	\centering
	\includegraphics[scale=0.9]{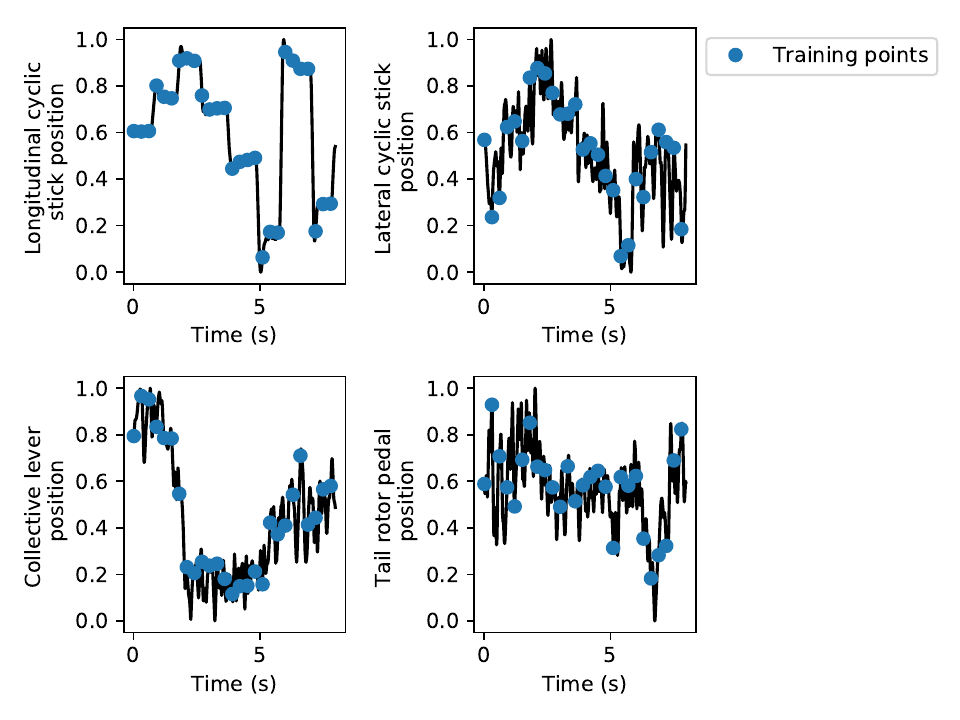}
    \caption{Pilot inputs during a maneuver of a Bo105 helicopter.}
    \label{fig:rotorcraft_training_inputs}
\end{figure}

\begin{figure}[H]
	\centering
	\includegraphics[scale=0.9]{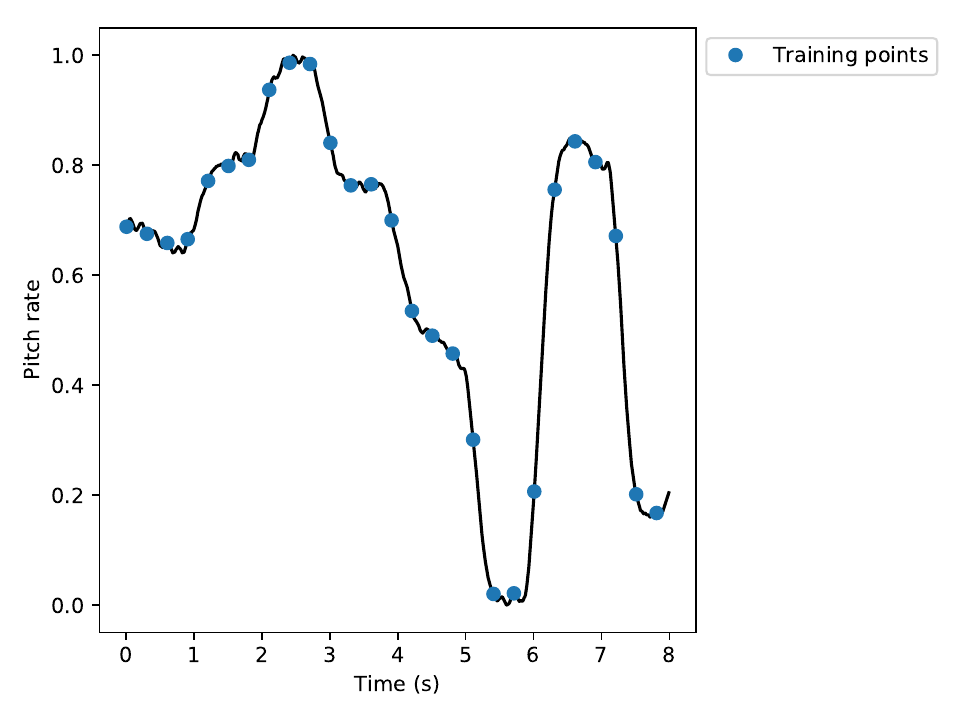}
    \caption{Pitch rate response during a maneuver of a Bo105 helicopter.}
    \label{fig:rotorcraft_training_outputs}
\end{figure}

Figure \ref{fig:rotorcraft_training_inputs} shows the (normalised) pilot inputs during a particular maneuver of a Bo105 helicopter while Figure \ref{fig:rotorcraft_training_outputs} shows the corresponding (normalised) pitch rate of the rotorcraft. We note here that, in the current example, we are investigating a model that is designed to predict \emph{on-axis} responses only - the main pilot input during this maneuver is longitudinal stick position and, as such, the model is designed to predict the rotorcraft pitch rate. \\

Defining $y_n$ as the rotorcraft's pitch rate at the $n$th time step then, following on from the work in \cite{jackson2019predicting}, predictions of $y_n$ are made according to a model of the form:

\begin{equation}
    y_n = f(y_{n-1}, \delta_n^x, \delta_n^y, \delta_n^p, \delta_n^o)
    \label{eq:flight_model}
\end{equation}
where $\delta^x$ is the longitudinal stick position, $\delta^y$ is the lateral stick position, $\delta^p$ is the tail rotor position and $\delta^o$ is the collective lever position. In the current example, a zero-mean Gaussian Process (GP) is used to model the relationship in equation (\ref{eq:flight_model}). The GP has a squared exponential kernel and employs a Gaussian likelihood, where it assumed that differences between the model's predictions and the measured values can be modelled as uncorrelated Gaussian noise with variance $\sigma^2$. The GP hyperparameters that require tuning are the length-scale of the kernel function, $l$, and the variance, $\sigma^2$, of the Gaussian noise. The training data are shown as blue points on Figures \ref{fig:rotorcraft_training_inputs} and \ref{fig:rotorcraft_training_outputs}. Here, the model is used to interpolate between the training points.  We note that the model being utilised can be classified as `GP-NARX' and, as such, its predictions have to be calculated in a Monte-Carlo fashion. This need arises from the fact that the model's own (uncertain) predictions are, in subsequent iterations, used as inputs (see \cite{worden2014gaussian} for a more detailed discussion). \\

\begin{figure}[H]
	\centering
	\includegraphics[scale=0.9]{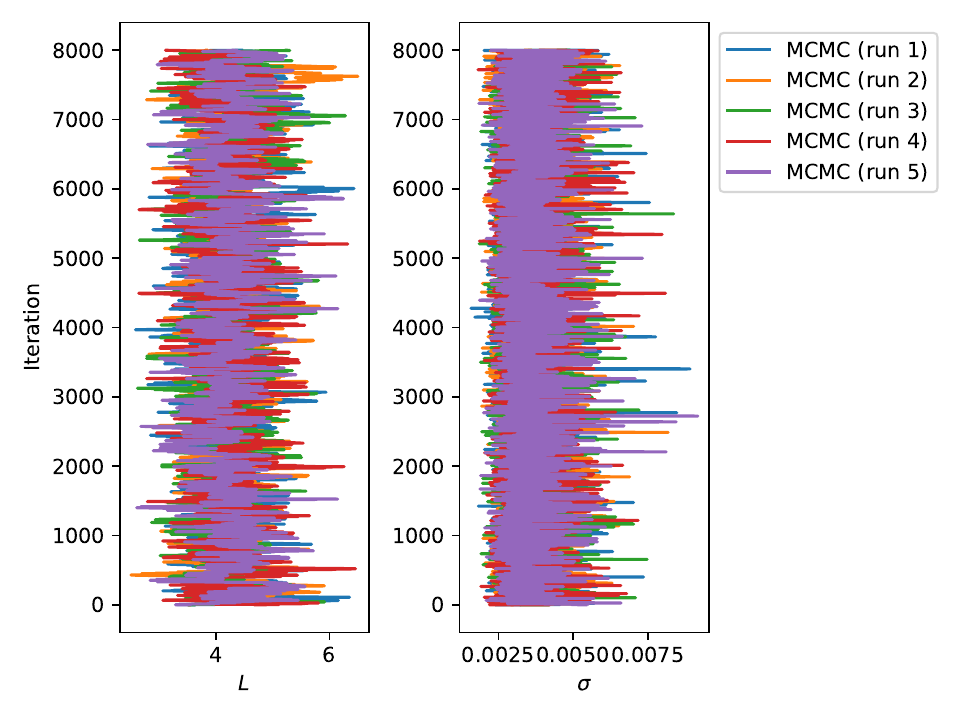}
    \caption{The results of 5 MCMC  short runs  for the Bayesian hyperparameter estimation of a Gaussian Process rotorcraft model.}
    \label{fig:rotorcraft_all_MCMC}
\end{figure}

\begin{figure}[H]
	\centering
	\includegraphics[scale=0.9]{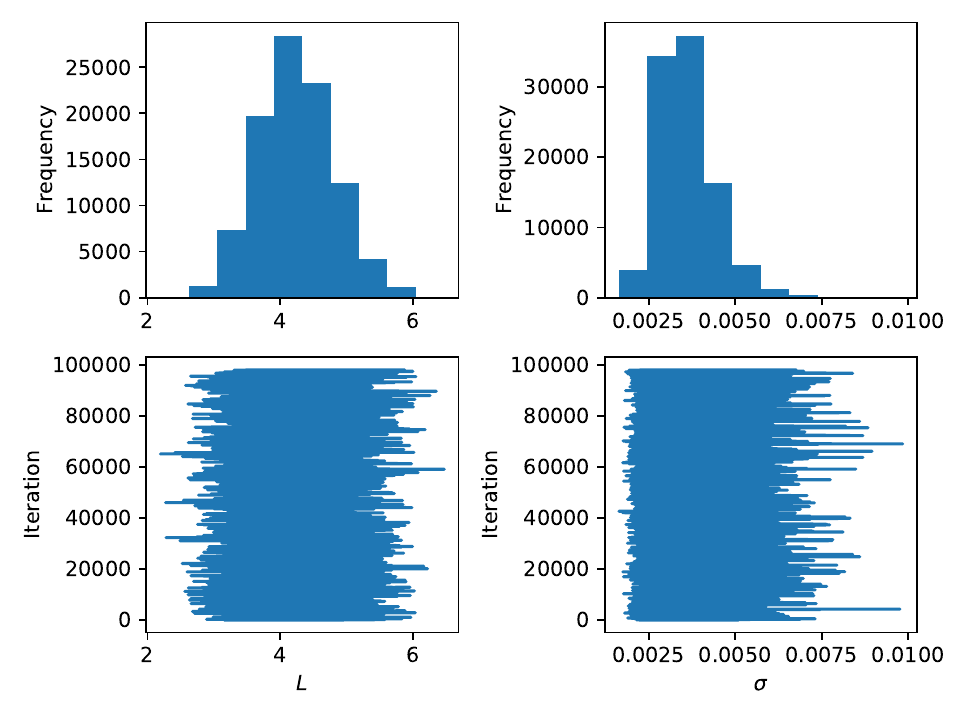}
    \caption{The results of a single MCMC  long run  for the Bayesian hyperparameter estimation of a Gaussian Process rotorcraft model.}
    \label{fig:rotorcraft_MCMC}
\end{figure}

In the following analysis, the aim is to establish whether uncertainties associated with the GP hyperparameters ($l$ and $\sigma$) have significant influence on the overall uncertainty associated with the model's predictions of pitch rate. To establish this, a Bayesian approach to the GP hyperparameter estimation is adopted where the following priors are used:

\begin{equation}
    p(l) = \text{Gamma}(l; a=1, s=1), \qquad p(\sigma) = \text{Gamma}(\sigma; a=1, s=0.01)
\end{equation}
where $a$ and $s$ are, respectively, the shape and scale parameters (as described in Scipy's \cite{2019arXiv190710121V} implementation of the Gamma distribution). To establish a `gold standard'  against  which we can compare our proposed SMC approach, MCMC was first used to sample from the GP hyperparameter posterior distribution. To this end, the Metropolis-Hastings algorithm was used to generate posterior samples (where the first 2,000 were then considered `burn-in' and removed accordingly). To provide a visual confirmation of MCMC convergence, 5  `short runs' (of 10,000 samples each) were conducted -  the resulting Markov chains are shown in Figure \ref{fig:rotorcraft_all_MCMC}.  A single MCMC `long run', of 100,000 samples, was then conducted.  The histograms associated with the MCMC long run are shown in Figure \ref{fig:rotorcraft_MCMC}. In the following, results realised from Figure \ref{fig:rotorcraft_MCMC} are used to validate the outputs of the proposed SMC sampler. Each SMC sampler was run for $k=500$ iterations using $N=1000$ samples per iteration. \\

\begin{figure}[H]
	\centering
	\includegraphics[scale=0.9]{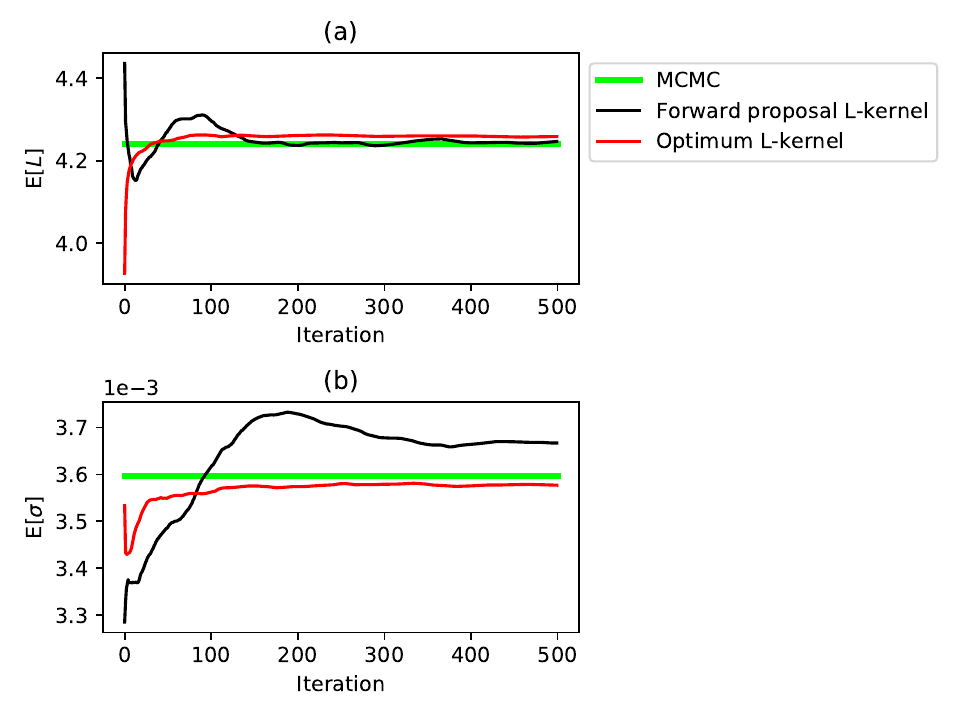}
    \caption{SMC sampler used to estimate the hyperparameter means for the rotorcraft problem described in Section \ref{sec:heli}. Black lines and red lines represent results obtained using a user defined L-kernel (equation (\ref{eq:q_L_1})) and a single-Gaussian approximately optimal L-kernel respectively. Green lines show the solution realised using MCMC.}
    \label{fig:rotorcraft_heli_means}
\end{figure}

\begin{figure}[H]
	\centering
	\includegraphics[scale=0.9]{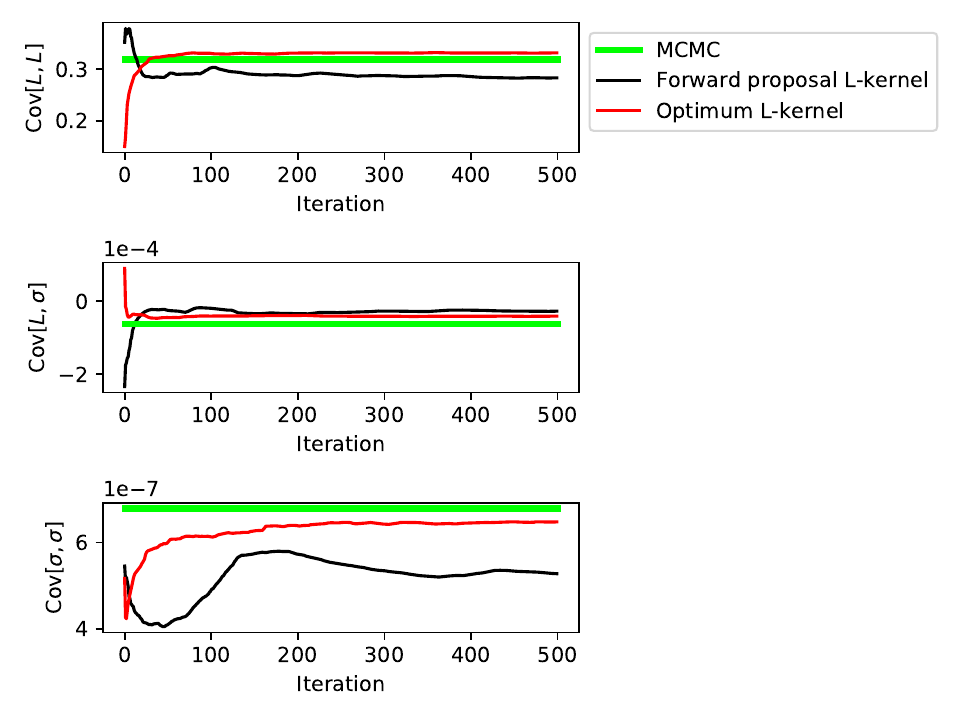}
    \caption{SMC sampler used to estimate the hyperparameter variance for the rotorcraft problem described in Section \ref{sec:heli}. Black lines and red lines represent results obtained using a user defined L-kernel (equation (\ref{eq:q_L_1})) and a single-Gaussian approximately optimal L-kernel respectively. Green lines show the solution realised using MCMC.}
    \label{fig:rotorcraft_heli_vars}
\end{figure}

\begin{figure}[H]
	\centering
	\includegraphics[scale=0.9]{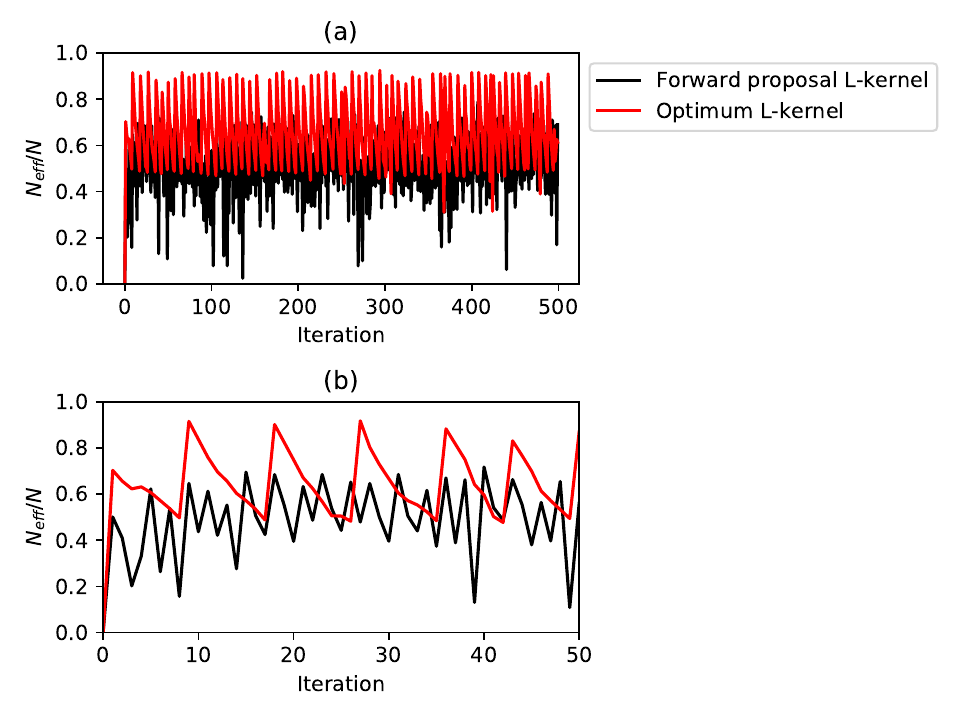}
    \caption{Effective sample size of SMC samplers applied to the rotorcraft problem described in Section \ref{sec:heli}. Black lines and red lines represent results obtained using a user defined L-kernel (equation (\ref{eq:q_L_1})) and a single-Gaussian approximately optimal L-kernel respectively. Panel (a) shows the effective sample size over the full 500 iterations, while panel (b) shows a zoomed-in portion of the results.}
    \label{fig:rotorcraft_heli_Neff}
\end{figure}

Figures \ref{fig:rotorcraft_heli_means} and \ref{fig:rotorcraft_heli_vars} show the estimated posterior hyperparameter means and covariances, using the recycling scheme described in Section \ref{sec:recycling}. It can be seen that, as with the previous examples, utilising the approximately optimal L-kernel has aided algorithm convergence.  Figure \ref{fig:rotorcraft_heli_Neff} shows that using the approximately optimal L-kernel has, again, helped the SMC sampler to maintain a higher effective sample size;  the forward proposal L-kernel and approximately optimal L-kernel implementations performed resampling 219 and 68 times respectively.  The sample variances achieved using both approaches are reported in Table \ref{table:help}.  We note here that 2-component approximations to the optimal L-kernel were also investigated but that no significant improvement in performance was observed (indicating that the uni-modality of the target distribution). The 2-component results are not reported here for visual clarity.   \\

\begin{figure}[H]
    \centering
    \includegraphics[scale=0.9]{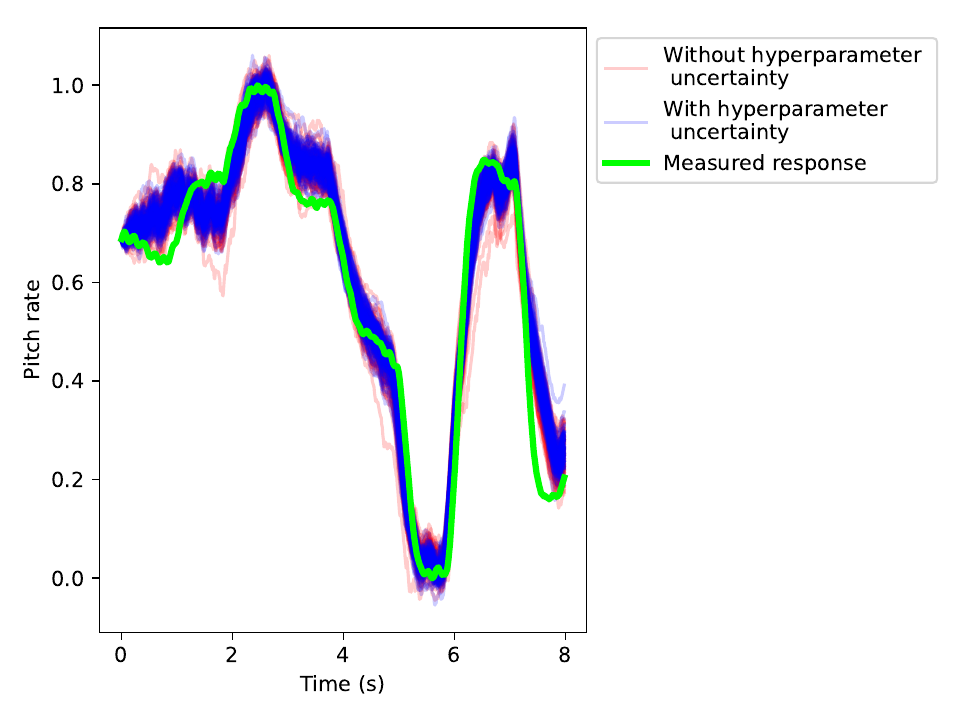}
    \caption{Predictions of rotorcraft pitch rate made by the GP-NARX model described in Section \ref{sec:heli}. Red lines indicate predictions made using the maximum posterior estimate of the GP hyperparameters. Blue lines indicate predictions made where hyperparameter uncertainty is accounted for via a Monte-Carlo approach. Red and blue lines are semi-transparent, such that darker regions indicate a higher density of Monte Carlo predictions. Green represents the measured response of the rotorcraft.}
    \label{fig:rotorcraft_heli_predictions}
\end{figure}

For the sake of completeness, Figure \ref{fig:rotorcraft_heli_predictions} shows Monte Carlo simulations of the model's predictions of pitch rate using a maximum posterior estimate of the GP hyperparameters (red lines) and accounting for hyperparameter uncertainty (blue lines). We note that the hyperparameter uncertainty does not appear to contribute significantly to the overall uncertainty of the modelling approach (a similar conclusion was reached, using MCMC, in \cite{jackson2018towards}).

\begin{table}[H]
    \centering
    \begin{tabular}{c|c|c}
         & Forward proposal L-kernel & Optimal L-kernel \\
        \hline
        $\E[L]$ & $8.420 \times 10^{-3}$ & $0.779 \times 10^{-3}$ \\
        $\E[\sigma]$ & $2.202 \times 10^{-8}$ & $0.167 \times 10^{-9}$ \\
        $\Cov[L, L]$ & $1.522 \times 10^{-3}$ & $0.344 \times 10^{-3}$ \\
        $\Cov[L, \sigma]$ & $1.631 \times 10^{-9}$ & $0.298 \times 10^{-9}$ \\
        $\Cov[\sigma, \sigma]$ & $1.352 \times 10^{-14}$ & $0.486 \times 10^{-14}$ \\
    \end{tabular}
    \caption{ Sample variance of the estimates realised for the example described in Section \ref{sec:heli}. }
    \label{table:help}
\end{table}

\section{Discussion}\label{sec:discussion}

The investigation detailed in the current paper highlights three considerations that must be made in the application of the proposed approximation approach for the optimal L-kernel. \\

Firstly, the approach relies on the approximation of the joint proposal distribution, $q(\x_{k-1}, \x_k)$, as a Gaussian mixture. The quality of this approximation will, of course, influence the success of the proposed approach. In particular, it has to be ensured that a sufficient number of samples are used at each iteration of the algorithm. The authors aim to investigate the influence of this requirement on higher dimensional problems as a topic of future work. \\

Secondly, the current work discusses an approximation for the optimal-L kernel, but it does not discuss the optimal choice of proposal distribution. The proposal distribution will, in a similar manner to MCMC, have significant influence over the efficiency of the algorithm. In short, if a poor proposal is selected, the algorithm will perform badly regardless of the choice of L-kernel. For future work, the authors aim to couple the proposed approach with a strategy for selecting good proposal distributions - this will likely utilise an `annealing-based' algorithm, that is common to many MCMC \cite{kirkpatrick1983optimization, szu1987fast, ingber1989very, marinari1992simulated, geyer1995annealing, ching2007transitional, green2015bayesian} and SMC \cite{zhou2013sequential, wan2011bayesian, bernardo2011free, nguyen2015efficient, svensson2015marginalizing} implementations. This work may exploit the fact that SMC samplers are not restricted by the condition known as `detailed balance' and can therefore potentially utilise a broader range of proposal distributions relative to MCMC. These proposal distributions may, for example, may incorporate \emph{memory} i.e., they may be formed using information realised from multiple iterations of previously generated samples. \\

Thirdly, we note that, throughout the current paper, the use of an approximately optimal L-kernel prevented relatively large drops in effective sample size and helped reduce the number of times resampling was required. We emphasise that this feature of the approach is particularly important regarding problems with multiple modes as, in such cases, it is possible (if improbable) that resampling will lead to the removal of samples from a particular mode. \\

\section{Conclusions}\label{sec:conclusions}
Markov chain Monte Carlo (MCMC) has become a key tool in the treatment of intractable Bayesian inference problems. Sequential Monte Carlo (SMC) samplers, however, provide an alternative to MCMC that is easier to parallelise and which has several `tuning parameters' that are not available to MCMC. The current paper investigates an implementation strategy for one of these tuning parameters - the SMC sampler L-kernel. Through the proposed strategy, a number of case studies are used to illustrate how the efficiency of the sampler can be improved. Specifically, for the examples shown in the current paper, the use of an approximately optimum L-kernel has reduced the sample variance of the SMC estimates by up to 99 \% while also reducing the number of times that resampling was required by between 65 \% and 70 \%. \\

\section*{Acknowledgements}
The authors gratefully acknowledge the Engineering and Physical Sciences Research Council, who funded this work through the grant `Big Hypotheses: a Fully Parallelised Bayesian Inference Solution' (EP/R018537/1). 

\bibliography{bibliography}
\bibliographystyle{unsrt}

\end{document}